\documentclass[journal,twocolumn]{IEEEtran}   
\usepackage{graphicx}
\usepackage{epstopdf}
\usepackage{amssymb}
\usepackage{amsfonts}
\usepackage{amsmath}
\usepackage{booktabs}
\usepackage{algorithm}
\usepackage{color}
\usepackage{algorithmic}
\usepackage{subeqnarray}
\usepackage{cases}
\usepackage{bm}
\usepackage{subfigure,amsmath,amssymb,cite}
\usepackage{stfloats}
\usepackage{booktabs}
\usepackage{float}
\ifCLASSINFOpdf
\else
\fi
\begin{document}

\title{Two Low-complexity Efficient Beamformers  for IRS-and-UAV-aided Directional Modulation Networks}

\author{~Yeqing Lin,~Feng Shu,~Yuxiang Zheng,~Jing Liu,~Rongen Dong,~Xun Chen,~Yue Wu,\\~Shihao Yan, and Jiangzhou Wang,\emph{ Fellow, IEEE}

\thanks{This work was supported in part by the National Natural Science Foundation of China (Nos.U22A2002, and 62071234), the Hainan Province Science and Technology Special Fund (ZDKJ2021022), and the Scientific Research Fund Project of Hainan University under Grant KYQD(ZR)-21008.(Corresponding authors: Feng Shu and Xun Chen)}
\thanks{Yeqing Lin,~Jing Liu, Rongen Dong,~Xun~Chen,~Yue Wu, and Feng Shu are with the School of Information and Communication Engineering, Hainan University, Haikou, 570228, China. (Email: shufeng0101@163.com, chenxun@hainan.edu.cn) }
\thanks{Feng Shu is with the School of Electronic and Optical Engineering, Nanjing University of Science and Technology, 210094, China. (Email: shufeng0101@163.com)}
\thanks{Jiangzhou Wang is with the School of Engineering, University of Kent, Canterbury CT2 7NT, U.K. (Email: j.z.wang@kent.ac.uk)}


}

\maketitle

\begin{abstract}
As the excellent tools for aiding communication, intelligent reflecting surface (IRS) and unmanned aerial vehicle (UAV) can extend the coverage area, remove blind area, and achieve a dramatic  rate improvement. In this paper, we improve the secrecy rate (SR) performance  at directional modulation (DM) networks using IRS and UAV in combination. To fully explore the benefits of IRS and UAV, two efficient methods  are proposed to enhance SR performance. The first approach  computes  the confidential message (CM) beamforming vector by maximizing the SR, and the signal-to-leakage-noise ratio (SLNR) method is used to optimize the IRS phase shift matrix, which is called  Max-SR-SLNR.  To reduce the computational complexity, the CM, artificial noise (AN) beamforming and IRS phase shift design are independently designed in the following methods. The CM beamforming vector is constructed based on maximum ratio transmission (MRT) criteria along the channel from Alice-to-IRS, the AN beamforming vector is  designed by null-space projection (NSP) on the remaining two channels, and phase shift matrix of IRS is directly given by phase alignment (PA) method. This method is called MRT-NSP-PA. Simulation results show that the SR performance of the Max-SR-SLNR method outperforms the MRT-NSP-PA method in the cases of small-scale and medium-scale IRSs, and the latter approaches the former in performance as IRS tends to lager-scale.
\end{abstract}
\begin{IEEEkeywords}
Intelligent reflecting surface, unmanned aerial vehicle, directional modulation, confidential message, artificial noise, secrecy rate.
\end{IEEEkeywords}

\section{Introduction}
\textcolor{blue}{Due to the advantages of mobile controllability, on-demand placement, and low cost, unmanned aerial vehicles (UAVs) have gained wide attention and application in industry and academia in \cite{xiao2016enabling,azari2020uav}. With lower manufacturing costs, miniaturization and improved performance of UAVs, many new applications have emerged for UAVs in civilian and commercial applications, such as cargo transportation, emergency search and rescue, and communication relay in \cite{zeng2016wireless}. In addition, UAVs have a high potential for having line-of-sight (LoS) air-to-ground communication links due to their mobile controllability and on-demand placement flexibility in \cite{zhao2020efficiency,wen20223}.}

However, the broadcast nature of wireless communications makes information vulnerable to interception or eavesdropping by unauthorized users\cite{broadcast1978,Mukherjee2014,Yangnan2013,hamamreh2018classifications,yang2015artificial}. Therefore, in the recent years,  the physical layer security (PLS) of wireless communication has attracted extensive attention. Moreover, PLS will  potentially become one of the key techniques of future sixth generation (6G) mobile communications. The main solution to the traditional PLS problem is to  prevent illegal users with a computational NP-complete difficulty from breaking the confidential message (CM) by using the encryption mechanism in the upper-layer protocol stack\cite{Diffie1976}.

Nevertheless, encryption technology that relies on encryption algorithms has security risks due to its conditional security. Therefore,
a keyless PLS security eavesdropping channel wire-tap model was proposed by Wyner\cite{wiretap1975}, secure communication can be achieved without relying on encryption technology. In \cite{Sarkar2010, khisti2008secure}, the authors proposed to use the artificial noise (AN) to effectively enhance the legitimate channels and weaken the eavesdropping channels in order to realize the wireless network security.
In \cite{Juying2018}, the authors made an investigation of  PLS with  multiple single-antenna eavesdroppers in millimeter wave channel. Here,  two transmission schemes, maximum ratio transmission (MRT) and  maximizing the security throughput under the constraint of security interruption probability,  were proposed by utilizing the specific propagation characteristics of mmWave.


As an effective physical layer transmission technology suitable for the LoS propagation channels, directional modulation (DM) attracts increasingly research attentions from both industry and academic world in \cite{Wuxiaoming2017,Hujinsong2016,Binqiu2019}. \textcolor{blue}{Considering that the air-to-ground and ground-to-air channels of UAV are mainly dominated by LoS channels, DM technology can be perfectly applied in UAV trunking network to significantly improve the secure performance of wireless network.} In \cite{mamaghani2019performance}, the authors studied the secure simultaneous wireless information and power transfer (SWIPT) problem for UAV relay networks, where the UAV operates in amplified forwarding (AF) mode, derive the connection probability, security outage probability, and instantaneous secrecy rate (SR) of the system, and analyze the asymptotic SR performance.  In secure SWIPT transmission network for UAV in millimeter wave scenarios \cite{sun2019secure}, the number and location of eavesdroppers obey the Poisson point process, and the UAV forwarded the CM while charging the legitimate user, jointly designed the launch power of the UAV and the placement position.
Adding AN to the DM network is an extraordinary effective way \cite{shu2018artificial}.
In \cite{Wuxiaoming2018Secure},  phase alignment (PA), AN, and random subcarrier selection based on orthogonal frequency division multiplexing were combined to achieve  a precise secure wireless communications.

Compared to relay \cite{Lichunguo2009}, intelligent reflecting surface (IRS) has the following main advantages: low-power consumption, low-cost and easy to realize large-scale or even ultra-large-scale. Thus, IRS is becoming an extremely hot research topic. By adjusting its phase shift matrix,  IRS may intelligently control and change wireless environment to improve system spectral efficiency and energy efficiency \cite{Wuqingqing2019,dong2022performance,Tangwankai2020,WuqingqingIntelligent,hum2013reconfigurable}.
IRS can be applied to a diverse variety of communication areas, such as multiple input multiple output (MIMO) \cite{HanYitao2022,Pancunhua2020}, relay\cite{Wangxuehui},  covert communication\cite{Zhouxiaobo}, SWIPT\cite{ShiWeipin2020}, spatial modulation networks \cite{Yanglili2022} and UAV network \cite{Pangxiaowei2021IRS}.
In \cite{Pancunhua2020}, adjusting the phase of IRS  was to mitigate the interference at cell-edge, and the performance of cell-edge user  was improved. In an IRS-assisted multi-antenna relay network \cite{Wangxuehui},  an alternating iterative structure was presented to jointly optimize the beamforming and the phase shift of the IRS to harvest a substantial rate performance gain. In IRS-aided covert communication \cite{Zhouxiaobo}, penalty successive convex approximation algorithm and a low-complexity two-stage algorithm to jointly design the reflection coefficient of IRS and the transmission power of Alice was proposed. More importantly, the authors proved the existence of perfect covertness under perfect channel state information (CSI).
In IRS-aided UAVs network, the average SR was improved by jointly optimizing the design of beamforming, joint trajectory and phase shift of IRS \cite{Pangxiaowei2021IRS}.


Combining  IRS and  DM  can make a dramatic rate performance improvement \cite{Hujingsong2020,chen2022artificial,LiJiayu2021,Dongrongen_receive}.  In \cite{Hujingsong2020}, using multipath channel, a single CM signal was transmitted from Alice to Bob using two symbolic time slots, where the IRS phase matrix was aligned with the direct and cascade paths, respectively. Although the method in \cite{Hujingsong2020} has low complexity, the transfer of information in two time slots leads to SR performance loss. In order to further enhance the SR \cite{chen2022artificial}, a semi-definite relaxation (SDR) method is used to optimize the CM beamforming, AN beamforming, and IRS phase shift matrices alternately with the goal of maximizing the SR. In traditional DM networks, only one bitstream can be transmitted between  base station and  user, even with multiple antennas. With the help of IRS, is it possible to achieve a multiple stream transmission via  controlled multipath  in the LoS channel. In \cite{LiJiayu2021}, with the aid of IRS, the DM can achieve two independent CM streams from Alice to Bob under a multipath channel. To investigate the impact of optimizing the receive beamforming (RBF) vector on the performance of IRS-aided DM system, two alternately optimizing methods of jointly designing RBF vectors and IRS phase shift matrix were proposed to maximize the receive power sum in \cite{Dongrongen_receive}.

 However, the proposed  method in \cite{chen2022artificial} is of high computational complexity with a high-rate-performance, and the proposed scheme is of a low spectral-efficiency with a large rate performance loss due to the fact two-symbol-period only transmits one symbol \cite{Hujingsong2020}. \textcolor{blue}{In practical applications, IRS needs to be fixed at a specific location. Therefore, IRS can be installed on drone to provide a more reliable and secure connectivity. The UAV-assisted IRS device may be viewed as a reflecting relay. For example, due to severe shadow fading in urban buildings or mountainous terrain, or due to damage to communications infrastructure caused by natural disasters. On-demand wireless systems using low-altitude UAVs are typically deployed faster, more flexible for reconfiguration, and may have better communication channels due to the presence of strong LoS links. This may lead to a significant performance improvement than direct communication. Combining UAV, IRS and DM will fully explore their advantages: high position, passive reflection,and directive property to form an enhanced secure wireless transmission.} In this paper, we will propose two beamforming methods to \textcolor{blue}{enhance SR}, which will strike a good balance between performance and complexity. The main contributions of the paper are as follows:
\begin{enumerate}
  \item The IRS-and-UAV-assisted DM network is established to transmit a single CM stream by making full use of the advantages of DM and IRS to improve SR performance. To obtain a high performance SR, the CM beamforming vector is given by the rule of maximizing the SR (Max-SR), and the method of maximizing the signal-to-leakage-noise ratio (SLNR) in \cite{sadek2007leakage} is used to design the phase shift matrix of the IRS. An alternately iterative process is introduced between  the CM beamforming vector and the IRS phase shift matrices to further improve the SR performance. AN is independently projected on the null-space of Alice-IRS and Alice-Bob channels, maximizing interference with Eve through direct channel.  The iterative process is related to the initial value. Therefore, the method has high computational Complexity.
  \item To reduce the computational complexity of the first method, a maximum ratio transmission (MRT)-based method is proposed. Here, the CM and AN beamforming vectors are constructed by using MRT and MRT-NSP, respectively whereas the IRS phase shift matrix is designed by using phase alignment (PA). It is particularly noted that  the three optimization variables (OVs) are designed independently and the method is called MRT-NSP-PA. In addition, CM beamforming only is aligned to the Alice-to-IRS channel, ignoring the direct channel, etc. Therefore, the relationship between the direction of CM beamforming and the number of IRS elements is observed by simluation. By designing different MRT methods at different IRS scales, the SR performance of the MRT-NSP-PA method is improved at small-scale and medium-scale IRSs.
\end{enumerate}

The remainder is organized as follows. Section II shows the system model and SR problem, two methods and one inquiry are proposed in Section III. In Section IV, numerical simulations and related analysis are presented. In the end, section V draws our conclusions.

$Notations$: In this paper, bold lowercase and uppercase letters represent vectors and matrices, respectively.
{Signs  $(\cdot)^H$, $(\cdot)^{-1}$, $(\cdot)^{\dagger}$ and $\|\cdot\|$ denote the conjugate transpose operation, inverse operation, pseudo-inverse and 2-norm operation, respectively. The notation $\textbf{I}_N$ indicates the $N\times N$ identity matrix. The sign $\mathbb{E}\{\cdot\}$ expresses the expectation operation, and $\operatorname{diag}(\cdot)$ denotes the diagonal operator.
\section{system model}
\begin{figure}[h]
\centering
\includegraphics[width=0.460\textwidth,height=0.25\textheight]{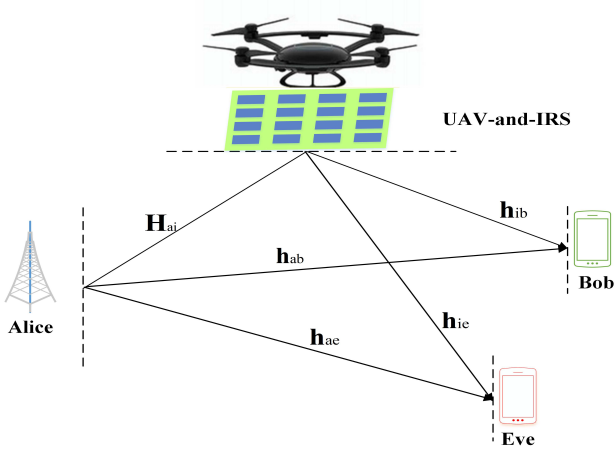}\\
\caption{System model diagram of \textcolor{red}{IRS-and-UAV-assisted} DM networks.}\label{systemmodel.eps}
\end{figure}
As shown in Fig.~\ref{systemmodel.eps}, an IRS-and-UAV-assisted DM network is sketched, \textcolor{red}{with IRS is installed on UAV, and UAV hovering at a suitable  height such that UAV can see Alice, Bob or both.} Alice is a transmitter equipped with $N_{a}$ antennas, both Bob and Eve are the legal and illegal receivers equipped with single antenna,respectively, and IRS has $N_{s}$ passive reflecting elements.

The baseband signal sent from Alice is given by
\begin{equation}
\mathbf{x}=\sqrt{\beta_{1} P_{t}}\mathbf{v}_{CM} s+\sqrt{\beta_{2} P_{t}} \mathbf{v}_{A N}{z},
\end{equation}
where $\beta_{1}$ and $\beta_{2}$ denote the power allocation (PA) factors for CM and AN with $\beta_{1}$ + $\beta_{2}$ = 1, $P_{t}$ stands for the transmit power, $\mathbf{v}_{CM}$ represents the precoding vector of the CM with $\mathbf{v}_{CM} \in \mathbb{C}^{N_{a} \times 1}$,  $\mathbf{v}_{AN}$ represents the precoding vector of the AN with $\mathbf{v}_{AN} \in \mathbb{C}^{N_{a} \times 1}$, $s$ represents the CM and with a constraint $\mathbb{E}\left[|s|^{2}\right]=1$, and $z$ is the AN with a constraint $\mathbb{E}\left[|z|^{2}\right]=1$.

The signal received at Bob is represented as
\begin{align}\label{Bob}
&y_{b}=\left(\sqrt{g_{a b}} \mathbf{h}_{a b}^{H}+\sqrt{g_{a i b}} \mathbf{h}_{i b}^{H} \bm{\Theta} \mathbf{H}_{ai}\right)\mathbf{x}+n_{b}\nonumber\\
&=\sqrt{g_{a b} \beta_{1} P_{t}} \mathbf{h}_{a b}^{H} \mathbf{v}_{CM} s+\sqrt{g_{a i b} \beta_{1} P_{t}} \mathbf{h}_{i b}^{H} \bm{\Theta} \mathbf{H}_{ai}\mathbf{v}_{CM} s+ \nonumber\\
&\sqrt{g_{a b} \beta_{2} P_{t}} \mathbf{h}_{a b}^{H} \mathbf{v}_{AN} z+\sqrt{g_{a i b} \beta_{2} P_{t}} \mathbf{h}_{i b}^{H} \bm{\Theta} \mathbf{H}_{ai}\mathbf{v}_{AN} z+n_{b},
\end{align}
where $\mathbf{h}_{ab} \in \mathbb{C}^{N_{a} \times 1}$ is the channel from Alice-to-Bob, $\mathbf{H}_{ai}=\mathbf{h}\left(\theta_{A I}^r\right) \mathbf{h}^H\left(\theta_{A I}^t\right) \in \mathbb{C}^{N_s \times N_a}$ represents the Alice-to-IRS channel,  $\mathbf{h}_{ib} \in \mathbb{C}^{N_{s} \times 1}$ is the channel from the IRS to Bob, $g_{ab}$ is the path loss coefficient between Alice and Bob, $g_{aib}$=$g_{ai}$$g_{ib}$ is the equivalent path loss coefficient of Alice-to-IRS channel and IRS-to-Bob channel, and $n_{b} \sim \mathcal{C} \mathcal{N}\left(0, \sigma_{b}^{2}\right)$ is the additive white Gaussian noise (AWGN) at Bob.
 The diagonal phase shifting matrix of the IRS is given by
 \begin{align}
\boldsymbol{\Theta}=\operatorname{diag}\left(e^{j \phi_{1}}, \cdots, e^{j \phi_{m}}, \cdots, e^{j \phi_{N_{s}}}\right),
\end{align}
where $\phi_{m}$ represents the phase shift of the m-th element on the IRS, and $\bm{\theta}^H= (e^{j \phi_{1}}, \cdots, e^{j \phi_{N}} )$.

The normalized channel vector can be expressed as
\begin{equation}
\mathbf{h}(\theta)=\frac{1}{\sqrt{N}}\left[e^{j 2 \pi \Phi_{1}\left(\theta\right)}, \cdots, e^{j 2 \pi \Phi_{n}\left(\theta\right)}, \cdots, e^{j 2 \pi \Phi_{N}\left(\theta\right)}\right]^{H},
\end{equation}
where the phase shift $\Phi_{n}(\theta)$ is defined as
\begin{align}\label{Phi_{n}}
\Phi_{n}(\theta)=-\frac{d}{\lambda}\left(n-\frac{N+1}{2}\right) \cos \theta, \quad n=1, \cdots, N,
\end{align}
where $d$ is the element spacing in the transmit antenna array, $\lambda$ is the wavelength, $n$ denotes the index of antenna, and $\theta$ is the direction angle of departure.

Similarly, the signal received at Eve can be given by
\begin{align}\label{Eve}
&y_{e}=\left(\sqrt{g_{a e}} \mathbf{h}_{a e}^{H}+\sqrt{g_{a i e}} \mathbf{h}_{i e}^{H} \bm{\Theta} \mathbf{H}_{ai}\right)\mathbf{x}+n_{e}\nonumber\\
&=\sqrt{g_{a e} \beta_{1} P_{t}} \mathbf{h}_{a e}^{H} \mathbf{v}_{CM} s+\sqrt{g_{a i e} \beta_{1} P_{t}} \mathbf{h}_{i b}^{H} \bm{\Theta} \mathbf{H}_{ai}\mathbf{v}_{CM} s+ \nonumber\\
&\sqrt{g_{a e} \beta_{2} P_{t}} \mathbf{h}_{a e}^{H} \mathbf{v}_{AN} z+\sqrt{g_{a i e} \beta_{2} P_{t}} \mathbf{h}_{i b}^{H} \bm{\Theta} \mathbf{H}_{ai}\mathbf{v}_{AN} z+n_{e},
\end{align}
where $\mathbf{h}_{ae}\in \mathbb{C}^{N_{a} \times 1}$ is the channel from Alice to Eve, $\mathbf{h}_{ie} \in \mathbb{C}^{N_{s} \times 1}$ represents the channel from IRS to Eve, $g_{ae}$ denotes the path loss coefficient between Alice and Eve, $g_{aie}$=$g_{ai}$$g_{ie}$ denotes the equivalent path loss coefficient of Alice-to-IRS channel and IRS-to-Eve channel, and $n_{e} \sim \mathcal{C} \mathcal{N}\left(0, \sigma_{e}^{2}\right)$ is the AWGN at Eve.

The signal received at IRS can be indicated as
\begin{equation}
y_{i}=\mathbf{H}_{ai}\mathbf{x}
=\mathbf{H}_{ai}(\sqrt{\beta_{1} P_{t}}\mathbf{v}_{CM} s+\sqrt{\beta_{2} P_{t}} \mathbf{v}_{A N}{z}),
\end{equation}

According to (\ref{Bob}), the signal-to-interference plus noise ratio (SINR) of Bob is given by
\begin{equation}
\gamma_{b}=\frac{\beta_{1} P_{t}\left|\sqrt{g_{a b}} \mathbf{h}_{ab}^{H} \mathbf{v}_{CM}+\sqrt{g_{a i b}} \mathbf{h}_{ib}^{H} \bm{\Theta} \mathbf{H}_{ai}\mathbf{v}_{CM}\right|^{2}}{\beta_{2} P_{t}\left|\sqrt{g_{a b}} \mathbf{h}_{ab}^{H} \mathbf{v}_{A N}+\sqrt{g_{a i b} } \mathbf{h}_{ib}^{H} \bm{\Theta} \mathbf{H}_{ai} \mathbf{v}_{A N}\right|^{2}+\sigma_{b}^{2}}.
\end{equation}
In terms of (\ref{Eve}), the SINR of Eve can expressed as
\begin{equation}
\gamma_{e}=\frac{\beta_{1} P_{t}\left|\sqrt{g_{a e}} \mathbf{h}_{ae}^{H} \mathbf{v}_{CM}+\sqrt{g_{a i e}} \mathbf{h}_{ie}^{H} \bm{\Theta} \mathbf{H}_{ai}\mathbf{v}_{CM} \right|^{2}}{\beta_{2} P_{t}\left|\sqrt{g_{ae}}\mathbf{h}_{ae}^{H} \mathbf{v}_{A N} +\sqrt{g_{a i e} } \mathbf{h}_{ie}^{H} \bm{\Theta} \mathbf{H}_{ai} \mathbf{v}_{AN}\right|^{2}+\sigma_{e}^{2}}.
\end{equation}
The corresponding rates of Bob and Eve are given by
\begin{align}
R_{b}=\log _{2}\left(1+\gamma_{b}\right),
\end{align}
and
\begin{align}
R_{e}=\log _{2}\left(1+\gamma_{e}\right),
\end{align}
respectively, which the calculation of secrecy rate is as follows
\begin{equation}\label{SR}
R_{s}=\left[R_{b}-R_{e}\right]^{+}=\log _{2}\left(\frac{1+\gamma_{b}}{1+\gamma_{e}}\right),
\end{equation}
where $[x]^{+} \triangleq \max \{0, x\}$.

\section{Proposed beamforming methods}
In this section, two beamforming methods, called Max-SR-SLNR and MRT-NSP-PA, are proposed in IRS-and-UAV-assisted DM networks.  The SR performance of the two proposed methods improves by about 30\% over the existing method \cite{Hujingsong2020}, and the complexity of the latter is two orders of magnitude lower than the former under the large-scale IRS. To further improve the SR performance of the MRT-NSP-PA, the relationship between the direction of CM beamforming and the number of IRS elements is explored.

\subsection{Proposed Max-SR-SLNR}

First, we optimize the AN beamforming vector, which is independent of $\bm{\theta}$ and $\mathbf{v}_{CM}$.   Alice projects the $\mathbf{v}_{AN}$ on the null space of Alice-to-Bob channel and Alice-to-IRS channel, and  maximize the received AN power along Alice-to-Eve direct channel at Eve.
The optimization problem of $\mathbf{v}_{AN}$ is given by
\begin{subequations}\label{vAN}
\begin{align}
\max _{\mathbf{v}_{AN}}  \quad& \mathbf{v}_{AN}^{H} \mathbf{h}_{ae}\mathbf{h}_{ae}^{H} \mathbf{v}_{AN} \\
\text { s.t. } \quad& \left( \mathbf{h}_{ab} \quad \mathbf{H}_{a i}^{H}\right)^{H} \mathbf{v}_{AN}=\mathbf{0} ,\quad\mathbf{v}_{AN}^{H} \mathbf{v}_{AN}=1.
\end{align}
\end{subequations}
Let us define
\begin{align}
\mathbf{P} = \left( \mathbf{h}_{ab} \quad \mathbf{H}_{ai}^{H}\right)^{H}.
\end{align}
According to the first constraint of (\ref{vAN}), $\mathbf{v}_{AN}$ can be rewritten as
\begin{align}
\mathbf{v}_{AN}=\left[\mathbf{I}_{N_{a}}-\mathbf{P}^{H}\left(\mathbf{P} \mathbf{P}^{H}\right)^{\dagger} \mathbf{P}\right] \mathbf{u}_{AN},
\end{align}
where $\mathbf{u}_{AN}$ is a new optimization variable with $\mathbf{u}_{AN}^{H} \mathbf{u}_{AN}=1$.
Let us define
\begin{align}
\mathbf{T}= \left[\mathbf{I}_{N}-\mathbf{P}^{H}\left(\mathbf{P} \mathbf{P}^{H}\right)^{\dagger} \mathbf{P}\right].
\end{align}
Therefore, (\ref{vAN}) can be simplified as
\begin{subequations}
\begin{align}
\max _{\mathbf{u}_{AN}}  \quad& \mathbf{u}_{AN}^{H}\mathbf{T}^{H} \mathbf{h}_{ae}\mathbf{h}_{ae}^{H}\mathbf{T} \mathbf{u}_{AN} \\
\text { s.t. } \quad&\mathbf{u}_{AN}^{H} \mathbf{u}_{AN}=1.
\end{align}
\end{subequations}
Since the  matrix $\mathbf{T}$ is a matrix of rank-one, $\mathbf{v}_{AN}$ can be expressed as
\begin{equation}
 \mathbf{v}_{AN}=\frac{\mathbf{T}_{-ae}  \mathbf{h}_{ae}}{\left\|\mathbf{T}_{-ae}  \mathbf{h}_{ae}\right\|}.
\end{equation}

Next, we design the alternating iterative optimization problem with two variables, $\mathbf{v}_{CM}$ and $\bm{\theta}$.
The optimization problem with the criterion of maximizing the SR can be expressed as
\begin{subequations}\label{SR.}
\begin{align}
&\max_{\mathbf{v}_{CM}, \boldsymbol{\Theta}} R_{s}\left(\mathbf{v}_{CM}, \boldsymbol{\Theta}\right) \\
&~~~\text { s.t. }~ \mathbf{v}_{CM}^{H} \mathbf{v}_{CM}=1, \bm{\theta}^H\bm{\theta}=N_{s}.
\end{align}
\end{subequations}
The SINR of Bob can be re-expressed as follows
\begin{align}
\gamma_{b}=\frac{\mathbf{v}_{CM}^{H} \mathbf{h}_{b1} \mathbf{h}_{b1}^H \mathbf{v}_{CM}}{\mathbf{v}_{AN}^{H} \mathbf{h}_{b2}\mathbf{h}_{b2}^H  \mathbf{v}_{A N}+\sigma_{b}^{2}},
 \end{align}
 where
 \begin{align}
\mathbf{h}_{b1}^H &=\left(\sqrt{\beta_{1} P_{t} g_{a b}} \mathbf{h}_{ab}^{H}+\sqrt{\beta_{1} P_{t} g_{a i b}} \mathbf{h}_{ib}^{H} \bm{\Theta} \mathbf{H}_{ai}\right),\nonumber\\
\mathbf{h}_{b2}^H &=\left(\sqrt{\beta_{2} P_{t} g_{a b}} \mathbf{h}_{ab}^{H}+\sqrt{\beta_{2} P_{t} g_{a i b}} \mathbf{h}_{ib}^{H} \bm{\Theta} \mathbf{H}_{ai}\right).
\end{align}
 Accordingly, the rate of Bob can be rewritten as
\begin{align}\label{Rb}
R_{b}=\log _{2}\left(1+\frac{\mathbf{v}_{CM}^{H} \mathbf{h}_{b1} \mathbf{h}_{b1}^H  \mathbf{v}_{CM}}{\mathbf{v}_{AN}^{H} \mathbf{h}_{b2} \mathbf{h}_{b2}^H  \mathbf{v}_{A N}+\sigma_{b}^{2}}\right).
\end{align}
Similarly, the SINR of Eve can be rewritten as
\begin{align}
\gamma_{e}=\frac{\mathbf{v}_{CM}^{H} \mathbf{h}_{e1} \mathbf{h}_{e1}^H  \mathbf{v}_{CM}}{\mathbf{v}_{AN}^{H} \mathbf{h}_{e2}\mathbf{h}_{e2}^H  \mathbf{v}_{A N}+\sigma_{e}^{2}},
\end{align}
where
\begin{align}
\mathbf{h}_{e1}^H =\left(\sqrt{\beta_{1} P_{t} g_{a e}} \mathbf{h}_{ae}^{H}+\sqrt{\beta_{1} P_{t} g_{a i e}} \mathbf{h}_{ie}^{H} \bm{\Theta} \mathbf{H}_{ai}\right),\nonumber\\
\mathbf{h}_{e2}^H =\left(\sqrt{\beta_{2} P_{t} g_{a e}} \mathbf{h}_{ae}^{H}+\sqrt{\beta_{2} P_{t} g_{a i e}} \mathbf{h}_{ie}^{H} \bm{\Theta} \mathbf{H}_{ai}\right).
\end{align}
Therefore, the rate of Eve can be expressed as follows
\begin{align}\label{Re}
R_{e}=\log _{2}\left(1+\frac{\mathbf{v}_{CM}^{H} \mathbf{h}_{e1} \mathbf{h}_{e1}^H  \mathbf{v}_{CM}}{\mathbf{v}_{AN}^{H} \mathbf{h}_{e2} \mathbf{h}_{e2}^H  \mathbf{v}_{A N}+\sigma_{e}^{2}}\right).
\end{align}

According to (\ref{Rb}) and (\ref{Re}), given $\bm{\Theta}$ and $\mathbf{v}_{AN}$, the optimization problem of (\ref{SR.}) can be converted into
\begin{subequations}
\begin{align}
&\max _{\mathbf{v}_{CM}} \frac{\mathbf{v}_{CM}^{H}\left((a+\sigma_{b}^2) \mathbf{I}_{N_{a}}+\mathbf{h}_{b1} \mathbf{h}_{b1}^H \right) \mathbf{v}_{CM}}  {\mathbf{v}_{CM}^{H}\left((b+\sigma_{e}^2) \mathbf{I}_{N_{a}}+\mathbf{h}_{e1} \mathbf{h}_{e1}^H \right) \mathbf{v}_{CM}}\\
&\text { s.t. } \quad \mathbf{v}_{CM}^{H} \mathbf{v}_{CM}=1,
\end{align}
\end{subequations}
where
$a=\mathbf{v}_{AN}^{H}\mathbf{h}_{b2}\mathbf{h}_{b2}^H \mathbf{v}_{AN}$, and $b=\mathbf{v}_{AN}^{H}\mathbf{h}_{e2}\mathbf{h}_{e2}^H \mathbf{v}_{AN}$
due to the fact that the logarithm function is a monotonically increasing function.

Accordingly, the Rayleigh-Ritz ratio theorem can be used, and $\mathbf{v}_{CM}$ is the eigenvector corresponding to the largest eigenvalue of the following formula
\begin{equation}\label{va}
{\left((b+\sigma_{e}^2) \mathbf{I}_{N_{a}}+\mathbf{h}_{e1} \mathbf{h}_{e1}^H \right)}^{-1}{\left((a+\sigma_{b}^2) \mathbf{I}_{N_{a}}+\mathbf{h}_{b1} \mathbf{h}_{b1}^H \right)}.
\end{equation}

The signal-to-leakage-noise ratio (SLNR) method is used to design the phase shift of IRS \cite{sadek2007leakage} as follows
\begin{subequations}\label{theta}
\begin{align}
\max _{\bm{\theta}}  \quad& \operatorname{SLNR}(\bm{\theta}) \\
\text { s.t. } \quad&\bm{\theta}^H\bm{\theta}=N_{s},
\end{align}
\end{subequations}
where the objective function of (\ref{theta}) is
\begin{equation}
\operatorname{SLNR}(\bm{\theta})=\frac{\mathbf{h}_{ib}^H \bm{\Theta}  \mathbf{H}_{ai} \mathbf{v}_{CM} \mathbf{v}_{CM}^H \mathbf{H}_{ai}^H \bm{\Theta} \mathbf{h}_{ib}}{\mathbf{h}_{i e}^H \bm{\Theta}  \mathbf{H}_{ai} \mathbf{v}_{CM} \mathbf{v}_{CM}^H \mathbf{H}_{ai}^H \bm{\Theta} \mathbf{h}_{ie}+\sigma_e^2}.
\end{equation}
Ones obtain
\begin{equation}
\operatorname{diag}\{\mathbf{a}\} \mathbf{b}=\operatorname{diag}\{\mathbf{b}\} \mathbf{a},
\end{equation}
where $\mathbf{a} \in \mathbb{C}^{N_{s} \times 1}$ and $\mathbf{b} \in \mathbb{C}^{N_{s} \times 1}$.
Therefore, the objective function of (\ref{theta}) can be expressed as
\begin{align}
\frac{\boldsymbol{\theta}^{H} \mathbf{A} \boldsymbol{\theta}}{\boldsymbol{\theta}^{H}\mathbf{B} \boldsymbol{\theta}}.
\end{align}
where $\mathbf{A}$ = $\operatorname{diag}\left(\mathbf{H}_{ai} \mathbf{v}_{CM}\right) \mathbf{h}_{i b} \mathbf{h}_{i b}^{H} \operatorname{diag}\left(\mathbf{H}_{ai} \mathbf{v}_{CM}\right)$, and $\mathbf{B}$ = $\operatorname{diag}\left(\mathbf{H}_{ai} \mathbf{v}_{CM}\right) \mathbf{h}_{i e} \mathbf{h}_{i e}^{H} \operatorname{diag}\left(\mathbf{H}_{ai} \mathbf{v}_{CM}\right)+\frac{\sigma_e^2}{N_s} \mathbf{I}_{N_{s}}$.

Accordingly, the Rayleigh-Ritz ratio theorem can be used, and $\bm{\theta}$ can be expressed as the eigenvector corresponding to the largest eigenvalue of the following formula
\begin{align}\label{u}
\mathbf{B}^{-1}\mathbf{A}.
\end{align}
Let us define the eigenvector corresponding to the largest eigenvalue of (\ref{u}) as $\mathbf{u}$. Since $\bm{\theta}$ has a constant mode constraint,  $\bm{\theta}$ can be expressed as
\begin{align}\label{Theta}
\bm{\theta}=e^{j \arg \mathbf{u}}.
\end{align}

Up to now, the CM beamforming vector, AN beamforming vector and IRS phase shift matrix have been designed. It is particularly noted that the AN beamforming  is independent of the CM beamforming and the IRS phase shift matrix, while the CM beamforming and the IRS phase shift matrix are mutually coupled. Therefore, it is necessary to alternately optimize  $\mathbf{v}_{CM}$ and $\bm{\theta}$ until $R_{s}^{(p)}-R_{s}^{(p-1)} \leq \epsilon$, where $p$ represents the number of iterations, and  the optimal $\mathbf{v}_{CM}$ and $\bm{\theta}$ can be iterated. The whole iterative process is listed in the following table.
\begin{algorithm}
\caption{Proposed Max-SR-SLNR method}
\begin{algorithmic}[1]
\STATE Set initial solution $\bm{\Theta}^{(0)}$, $\mathbf{v}_{CM}^{(0)}$ and $\mathbf{v}_{AN}$. Random multiple phases of $\bm{\theta}$, and calculate the initial $R_{s}^{(0)}$.
\STATE  Set $p$=0, threshold $\epsilon$..
\REPEAT
\STATE Given ($\bm{\Theta}^{(p)}$,$\mathbf{v}_{AN}$), according to (\ref{va}) to get $\mathbf{v}_{CM}^{(p+1)}$.
\STATE Given ($\mathbf{v}_{CM}^{(p+1)}$,$\mathbf{v}_{AN}$), according to (\ref{Theta}) to get $\bm{\Theta}^{(p+1)}$.
\STATE Compute $R_{s}^{(p+1)}$ using $\mathbf{v}_{CM}^{(p+1)}$,$\mathbf{v}_{AN}$ and $\bm{\Theta}^{(p+1)}$.
\STATE $p$=$p$+1;
\UNTIL $R_{s}^{p}-R_{s}^{p-1} \leq \epsilon$, and record the maximum SR value.
\end{algorithmic}
\end{algorithm}

The computational complexity of the proposed Max-SR-SLNR method is
\begin{align}
&\mathcal{O}(D_{1}(D_{2}(N_{s}^{3}+7N_{s}^{2}+8N_{a}N_{s}-2N_{s}-2+\nonumber\\
&2N_{a}^3+4N_{a}^2)+2N_{a}^{2}+ N_{a}-1))
\end{align}
float-point operations (FLOPs), where $D_{1}$ and $D_{2}$ represent the iterative numbers of optimization variables $\mathbf{v}_{CM}$ and $\bm{\theta}$.

\subsection{Proposed MRT-NSP-PA}
In the above subsection, the  iterative optimization process between  variables $\mathbf{v}_{CM}$ and $\bm{\theta}$  led to a high computational complexity. In order to reduce the complexity, a low-complexity MRT-NSP-PA method is proposed in which the three variables $\mathbf{v}_{CM}$, $\mathbf{v}_{AN}$ and $\bm{\theta}$ are designed independently in the following.

Let us define
\begin{align}
 \mathbf{h}_{ai}=\mathbf{h}^H\left(\theta_{A I}^t\right).
\end{align}
First, the MRT method is used to design $\mathbf{v}_{CM}$. Taking the transmit power limit into account,  the final  CM beamforming vector can be directly given by
\begin{equation}
\mathbf{v}_{CM}=\frac{\mathbf{h}_{ai}}{\left\|\mathbf{h}_{ai}\right\|}.
\end{equation}
In the same manner, the AN beamforming method based on MRT and NSP is
\begin{equation}
 \mathbf{v}_{AN}=\frac{\mathbf{T}_{-ae}  \mathbf{h}_{ae}}{\left\|\mathbf{T}_{-ae}  \mathbf{h}_{ae}\right\|}.
\end{equation}
Now, we design the IRS phase matrix $\bm{\theta}$, which  is fully  different from the former two vectors. The receive CM power via the cascaded path at Bob is equal to
\begin{align}\label{Pb}
P_{b}=\beta_{1} P_{t}g_{a i b} \mathbf{v}_{CM}^{H} \mathbf{H}_{ai}^{H} \boldsymbol{\Theta}^{H}
\mathbf{h}_{i b} \mathbf{h}_{i b}^{H} \boldsymbol{\Theta} \mathbf{H}_{ai} \mathbf{v}_{CM}.
\end{align}
(\ref{Pb}) can be rewritten as
\begin{align}
P_{b}=&\beta_{1} P_{t}g_{aib} \boldsymbol{\theta}^{H} \operatorname{diag}\left(\mathbf{H}_{ai} \mathbf{v}_{CM}\right) \mathbf{h}_{i b}\cdot\\ &\mathbf{h}_{i b}^{H} \operatorname{diag}\left(\mathbf{H}_{ai} \mathbf{v}_{CM}\right) \boldsymbol{\theta}.
\end{align}
To maximize the receive CM power along the cascaded  path from Alice to Bob via IRS at Bob, the PA method directly gives the value of $\bm{\theta}$ as follows
\begin{align}
\bm{\theta}=e^{-j \arg \left(\mathbf{h}_{ib}^{H} \operatorname{diag}\left(\mathbf{H}_{ai} \mathbf{v}_{CM}\right)\right)^{H}}.
\end{align}


The complexity of this algorithm is
\begin{align}
\mathcal{O}\left(2N_{s}^{2}+2N_{a}N_{s}-2N_{s}+4 N_{a}+2 N_{a}^{2}-2\right)
\end{align}
FLOPs.

In the above, the CM beamforming is only phase-aligned the Alice-to-IRS channel, ignoring the direct path in the desired user Bob, etc. In order to evaluate the impact of the CM beamforming direction on SR performance, we explore the relationship between the number of IRS elements and the direction of CM beamforming. Thus, the CM beamforming is allowed to rotate in the  angle range during $[0, \pi]$.
In this case, the direction of CM beamforming $\theta_{CM}$ is written as
\begin{align}
\theta_{CM}\in [0 ,\pi].
\end{align}
In what follows, we adopt three methods to design $\mathbf{v}_{CM}$ as follows
\begin{equation}
\mathbf{v}_{CM}=\frac{\mathbf{h}_{ab}}{\left\|\mathbf{h}_{ab}\right\|},
\end{equation}
\begin{equation}
\mathbf{v}_{CM}=\frac{(\mathbf{h}_{ai}+\mathbf{h}_{ab})}{\left\|(\mathbf{h}_{ai}+\mathbf{h}_{ab})\right\|},
\end{equation}
and
\begin{equation}
\mathbf{v}_{CM}=\frac{\mathbf{h}_{ai}}{\left\|\mathbf{h}_{ai}\right\|}.
\end{equation}

\section{Simulation and Discussion}
In this section, the numeral results to examine the performance of our proposed algorithms are provided. Simulation parameters are set as follows: $P_{s}$ = 30 dBm, $\sigma_{b}^{2}=\sigma_{e}^{2}$= -40dBm, $N_{a}$ = 16, the PA factor is set as $\beta_{1}$ = 0.8. The BS, the IRS, Bob and Eve are located at (0 m, 0 m, 0 m), (0 m, 39.9 m, 3.5 m), (0 m, 90 m, 0 m) and (0 m, 96.6 m, 29.4 m), respectively. The coordinates and departure angles of the two-dimensional plane are set as follows. The distances are set as $d_{ai}$ = 40 m, $d_{ab}$ = 90 m, and $d_{ae}$ = 100 m, respectively. The angles of departure of each channel are set as $\theta_{ai}$ = $17\pi/36$, $\theta_{ab}$ = $1\pi/2$ and $\theta_{ae}$ = $7\pi/12$, respectively. The terminal parameter $\epsilon$ of the Max-SR-SLNR method is $10^{-3}$.

In the following, our two proposed methods are compared with the two benchmark schemes below:
\begin{enumerate}
  \item $\textbf{No-IRS}$: All IRS phase shift values are set to 0, i.e., $\bm{\Theta}$=$\mathbf{0}_{N_{s}\times N_{s}}$.
  \item $\textbf{Random Phase}$: The IRS phase shift value takes on a random value, and each IRS phase shift value is randomly distributed within $[0 , 2\pi)$.
\end{enumerate}

\begin{figure}[htbp]
\centering
\includegraphics[width=0.50\textwidth]{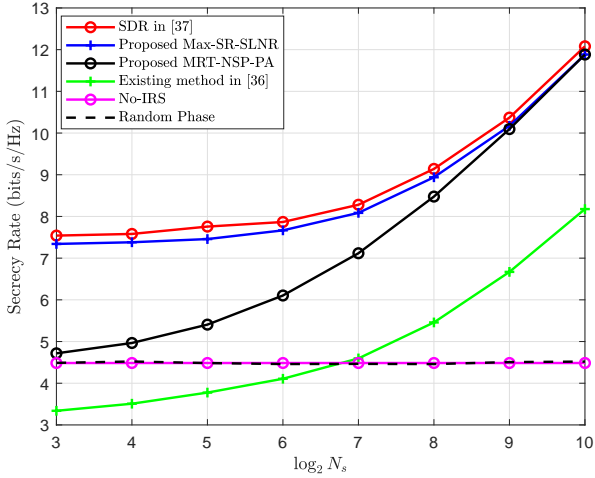}\\
\caption{Secrecy rate versus the number of IRS  elements}\label{SR_Nr.eps}
\end{figure}
Fig.~\ref{SR_Nr.eps}. shows that the SR versus the number of IRS elements for our proposed two methods with no IRS and random phase as performance benchmarks. The SR performance of the two proposed methods is much better than the cases of no IRS, random phase and existing method in \cite{Hujingsong2020}, and gradually grows with $N_{s}$. Their SR performance is ordered from the best to the worst as follows: SDR in \cite{chen2022artificial}, Max-SR-SLNR, MRT-NSP-PA, and existing method in \cite{Hujingsong2020}. The SR performance gap between Max-SR-SLNR and SDR in \cite{chen2022artificial} is small. The SR performance of the Max-SR-SLNR method is much better than that of the MRT-NSP-PA method when the IRS is  small to medium scale. For the case of large-scale, the latter approaches the Max-SR-SLNR in terms of SR.

\begin{figure}[htbp]
\centering
\includegraphics[width=0.50\textwidth]{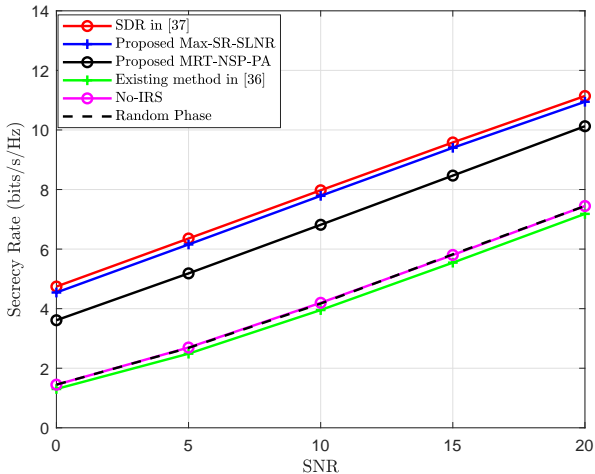}\\
\caption{Secrecy rate versus the SNR }\label{RSSNR.eps}
\end{figure}
Fig.~\ref{RSSNR.eps} plots the SR versus SNR for $N_{s}$=128. It can be seen that the SR performance of the six schemes increases with the increase of SNR. The SR performance of the proposed MRT-NSP-PA method and Max-SR-SLNR method is about 2 times those of the existing methods in \cite{Hujingsong2020}. When SNR=10dB, compared with the cases without IRS and random phase, SDR in \cite{chen2022artificial}, Max-SR-SLNR method and MRT-NSP-PA method achieve about  the SR improvements roughly: 47\%, 46\% and 34\%, respectively. Therefore, this means that optimizing the phase-shift matrix of IRS, CM beamforming, AN beamforming  can harvest  obvious performance gains.

Fig.~\ref{SRDAB.eps}. illustrates the SR versus the distance between Alice and Bob for $N_{s}$=100. It can be observed, as the distance between Alice and Bob increases, the SRs in all cases have the tendency of decreasing. This is because increasing the $d_{ab}$ will increase the path loss. It can be observed from the figure that the SR performance of the existing methods is only half of the SR performance of proposed two methods, thus demonstrating the advantages of our proposed methods.
\begin{figure}[htbp]
\centering
\includegraphics[width=0.50\textwidth]{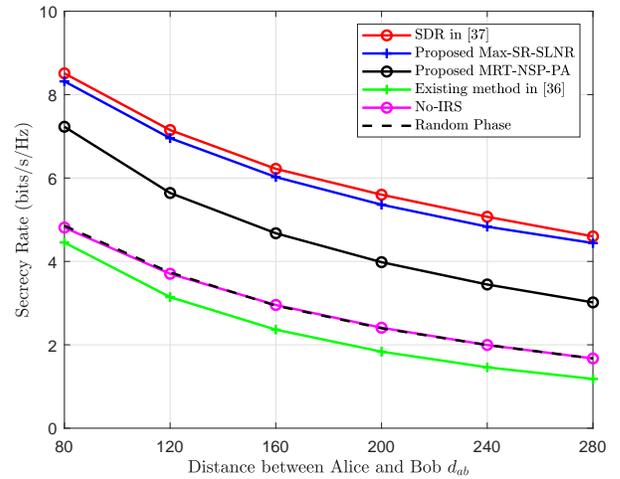}\\
\caption{Secrecy rate versus distance between Alice and Bob }\label{SRDAB.eps}
\end{figure}

\begin{figure*}
 \setlength{\abovecaptionskip}{-5pt}
 \setlength{\belowcaptionskip}{-10pt}
 \centering
 \begin{minipage}[htbp]{0.33\linewidth}
  \centering
  \includegraphics[width=2.56in]{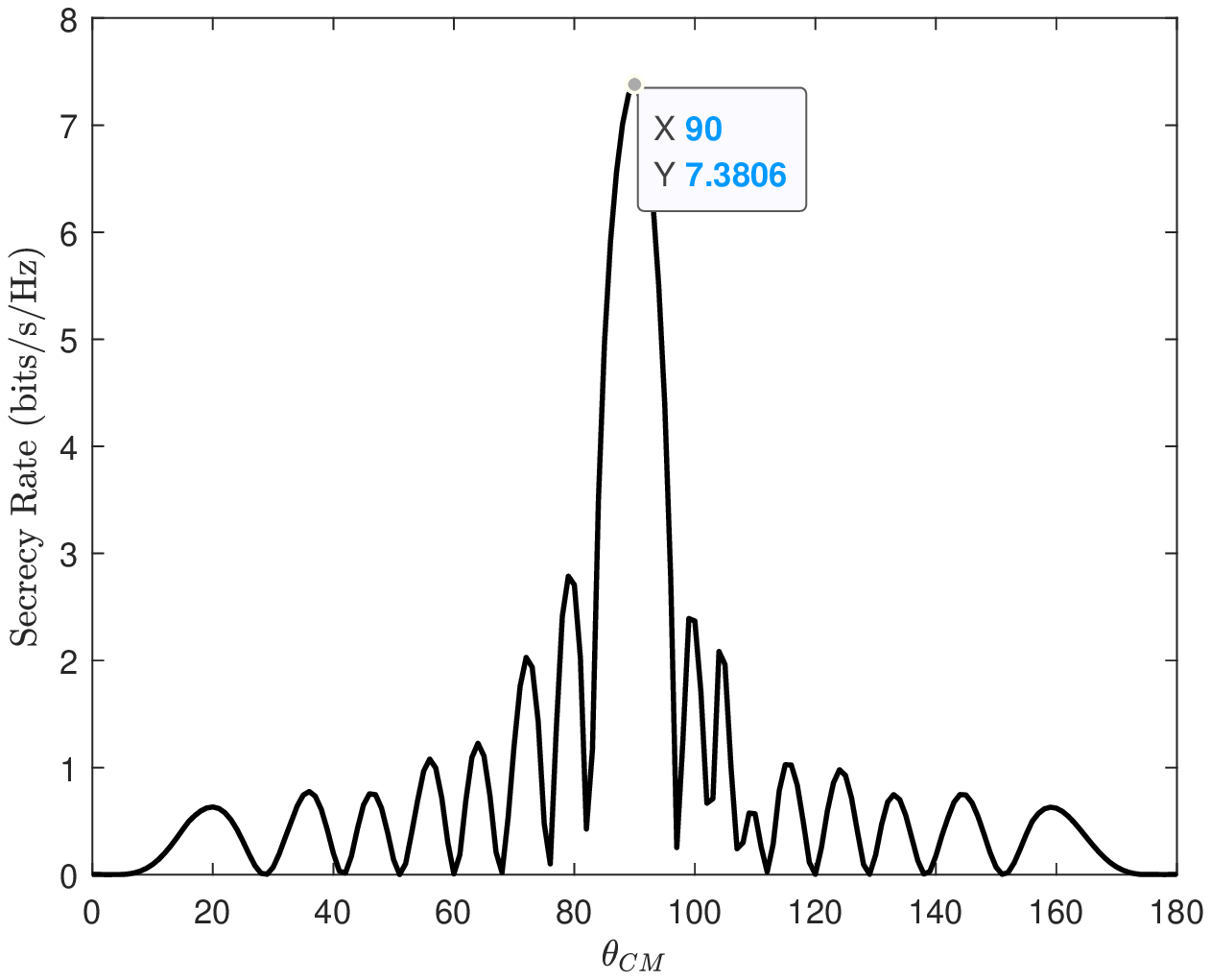}
  \caption{Secrecy rate versus  $\theta_{CM}$ }
 \end{minipage}%
 \begin{minipage}[htbp]{0.33\linewidth}
  \centering
  \includegraphics[width=2.56in]{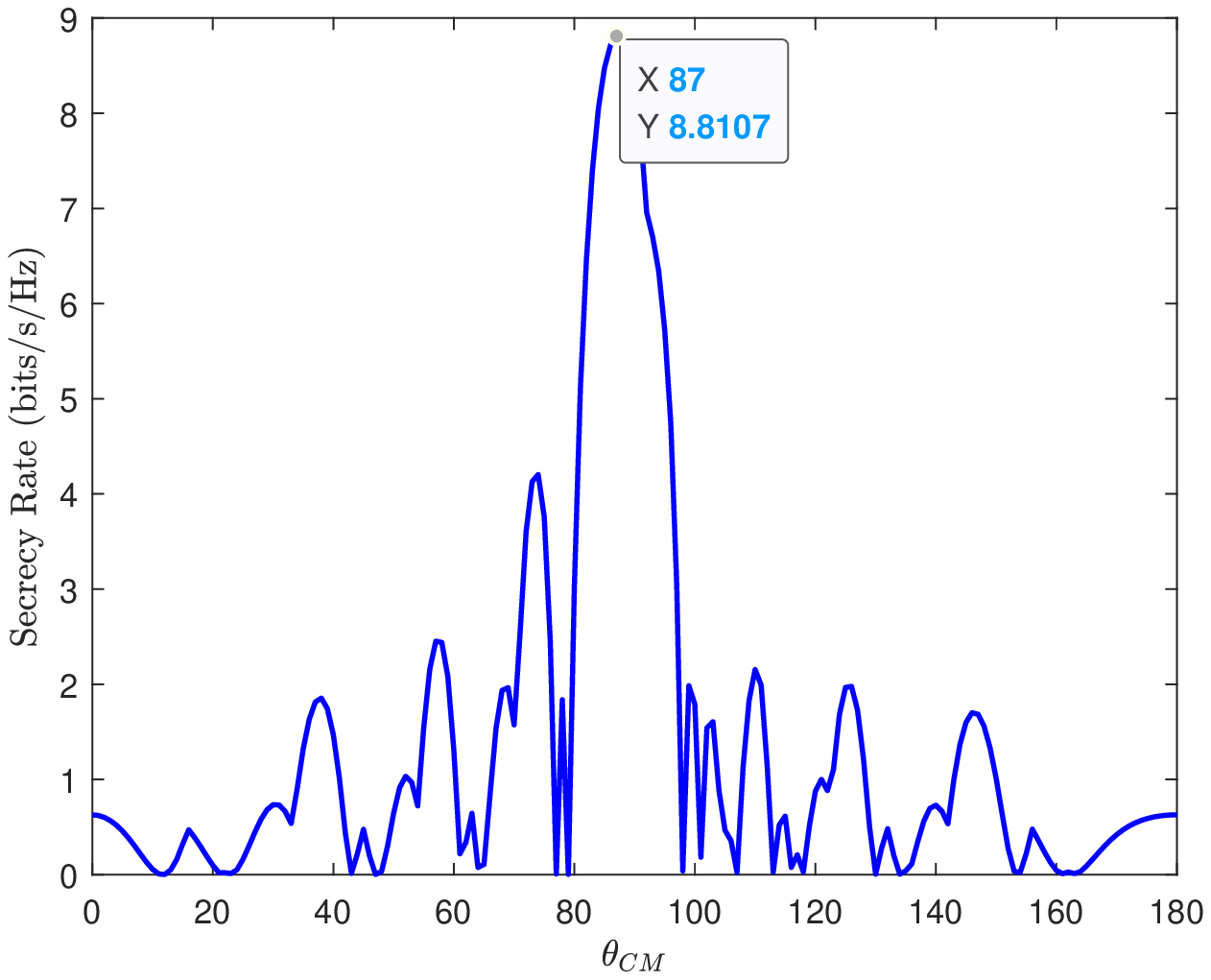}
  \caption{Secrecy rate versus  $\theta_{CM}$}
 \end{minipage}
 \begin{minipage}[htbp]{0.33\linewidth}
  \centering
  \includegraphics[width=2.56in]{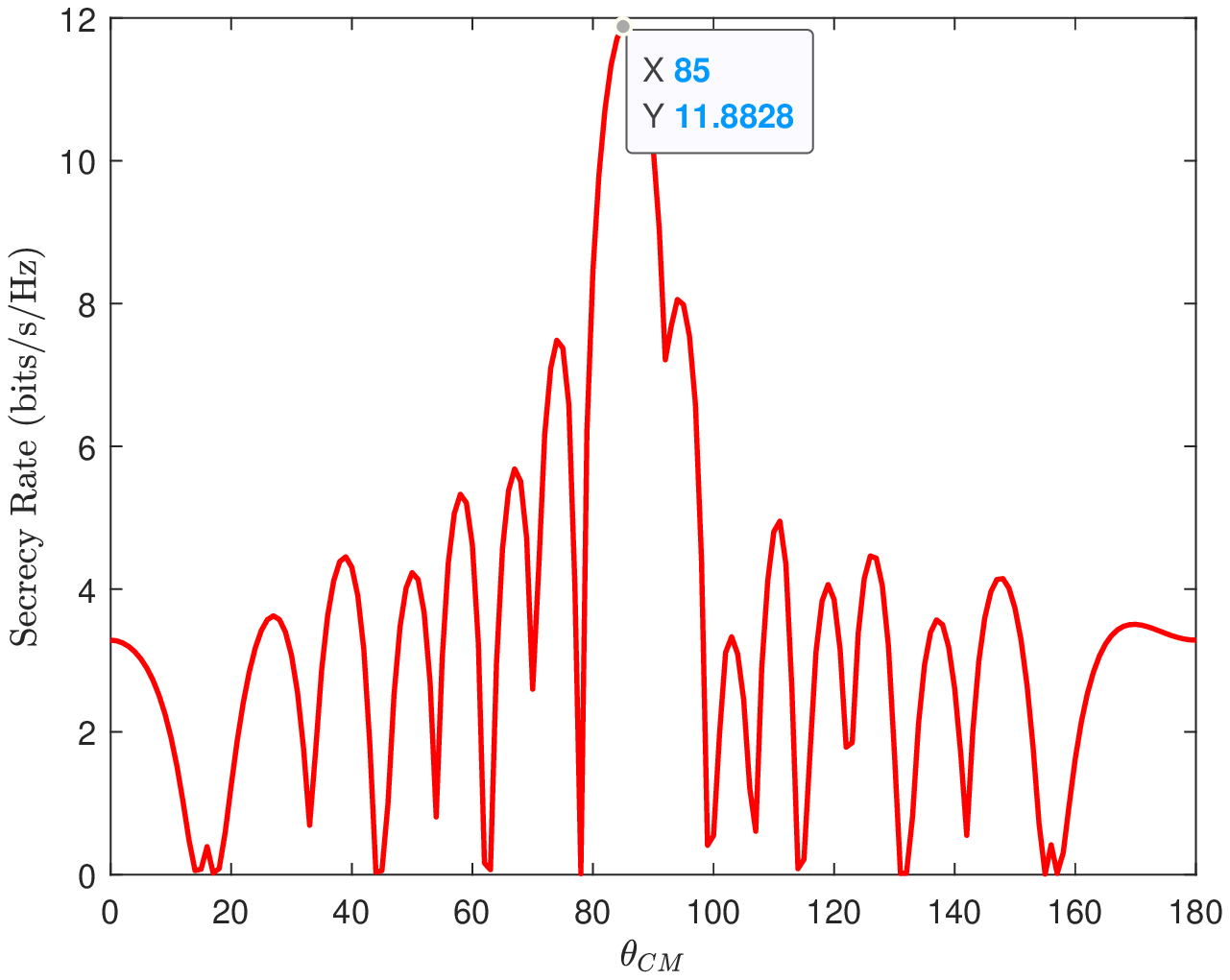}
  \caption{Secrecy rate versus  $\theta_{CM}$}
 \end{minipage}
\end{figure*}

Fig.~5.-Fig.~7. illustrate the SR versus directional angle of CM beamforming with different number of IRS elements as follows: 16, 256 and 1024. With the increase of the number of IRS elements, the directional angle of CM beamforming is constantly changing from $0$ to $\pi$. When $N_{s}$=16, the SR is the highest when the CM beamforming is transferred to the direct channel from Alice-to-Bob. When $N_{s}$=256, CM beamforming is directed to the middle of Alice-to-IRS channel and Alice-to-Bob channel. When $N_{s}$=1024, the SR is the highest when the CM beamforming is aimed at Alice-to-IRS channel. These results are mainly due to the fact that the Alice-to-Bob direct channel dominates  in the case of small-scale IRS whereas the cascaded channel via Alice, IRS and Bob dominates for the large-scale scenario.
\begin{figure}[htbp]
\centering
\includegraphics[width=0.50\textwidth]{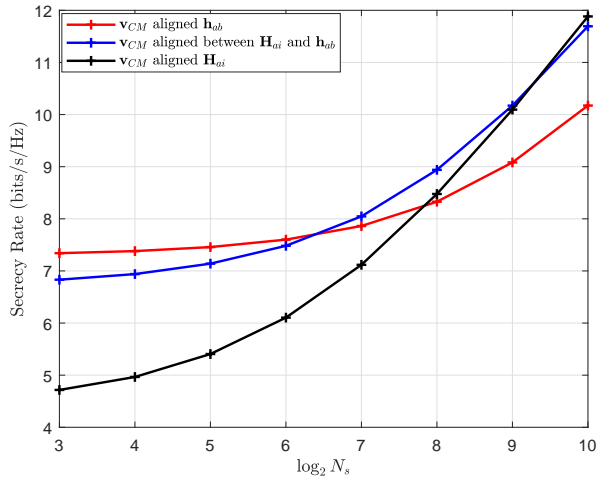}\\
\caption{Secrecy rate versus the number of IRS elements }\label{vcm.eps}
\end{figure}
\begin{figure}[htbp]
\centering
\includegraphics[width=0.50\textwidth]{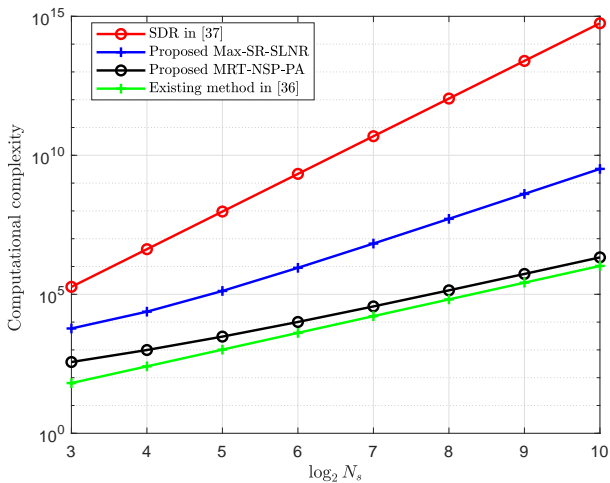}\\
\caption{Computational complexity versus the number of IRS elements }\label{flops.eps}
\end{figure}
Based on the inspiration of Figs. 5.-7., three different MRT methods for designing CM beamforming vectors are proposed. In Fig.~\ref{vcm.eps}., SR versus the number of IRS elements is plotted for different MRT methods. It can be seen from the figure that the three methods have different advantages under different numbers of IRS elements. When $N_{s}$ ranges from 8 to 64,   $\mathbf{v}_{CM}$ is aligned with the channel $\mathbf{h}_{ab}$  to achieve the best SR performance. When $N_{s}$ varies from 64 to 512, $\mathbf{v}_{CM}$ is aligned with the channel between $\mathbf{h}_{ab}$ and $\mathbf{h}_{ai}$  to achieve the best SR performance. And when $N_{s}$ changes from 512 to 1024 (i.e. under hyperscale), $\mathbf{v}_{CM}$ is aligned with the channel $\mathbf{h}_{ai}$  to achieve the best SR performance.

Fig.~\ref{flops.eps}. plots the computational complexity versus the number $N_s$ of IRS elements. From this figure, it is seen that there is an increasing order in complexity (FLOPs): existing method in  \cite{Hujingsong2020}  ($\mathcal{O}(N^2)$ FLOPs), MRT-NSP-PA ($\mathcal{O}(N^2)$ FLOPs), Max-SR-SLNR ($\mathcal{O}(N^3)$ FLOPs), and SDR in \cite{chen2022artificial} ($\mathcal{O}(N^{4.5})$ FLOPs).  Simulation results show that the complexity of the MRT-NSP-PA is at least one and two orders of magnitude lower than the Max-SR-SLNR for small-scale and large-scale IRS, respectively. Moreover, as the number of IRS elements increases, the complexity gradually and linearly increases. Thus, we conclude the proposed two methods  strike a good balance between SR performance and complexity.

\section{Conclusions}
In this paper, we have designed the beamforming of IRS-and-UAV-aided DM networks in order to fully exploit the SR performance benefit from IRS. Two beamforming methods, called Max-SR-SLNR and MRT-NSP-PA, were proposed. Simulation results showed that the two proposed methods can achieve an obvious SR performance gains over no-IRS, random phase, and existing method, especially in large-scale IRS. Moreover, the SR gains harvested by the proposed two methods grows gradually with the number of IRS elements increases. In the small-scale and medium-scale IRSs, the proposed Max-SR-SLNR method is better than the MRT-NSP-PA method in terms of SR and the latter approaches the former as the number of IRS elements goes to large-scale. However, the latter is at least one to two orders of magnitude lower than the former when the IRS size ranges from small to large.  In additional,  the proposed Max-SR-SLNR and MRT-NSP-PA have stricken a good balance between rate performance and complexity.

\ifCLASSOPTIONcaptionsoff
\newpage
\fi

\bibliographystyle{IEEEtran}
\bibliography{IEEEfull,reference}
\end{document}